\newcommand{\pbra}{\left\langle}
  \newcommand{\pket}{\right\rangle}
\newcommand{\bra}{\langle}
\newcommand{\ket}{\rangle}

\newcommand{\na}{\nabla}
\newcommand{\pa}{\partial}
\newcommand{\al}{\alpha}

\newcommand{\pder}[2]{\frac{\partial #1}{\partial  #2}}

\documentclass[12pt]{iopart}
\usepackage{graphicx}% Include figure files
  
\begin{document}

\title{Electroviscous effects of simple electrolytes under shear}

\author{Hirofumi Wada 
\footnote[3]{To
whom correspondence should be addressed (wada@daisy.phys.s.u-tokyo.ac.jp)}
}

\address{Department of Physics, University of Tokyo, Hongo, Tokyo, 113-0033,
Japan}

\begin{abstract}
On the basis of a hydrodynamical model analogous to that in critical fluids,  
we investigate the influences of shear flow upon the electrostatic
contribution to the viscosity of binary electrolyte solutions in the Debye-H\"{u}ckel
approximation.
Within the linear-response theory, we reproduce the classical limiting law that 
the excess viscosity is proportional to the square root of the concentration of
the electrolyte.
We also extend this result for finite shear. 
An analytic expression of the anisotropic structure factor of the charge density 
under shear is obtained, and its deformation at large shear rates is discussed.  
A non-Newtonian effect caused by deformations of the 
ionic atmosphere is also elucidated for $\tau_D\dot{\gamma}>1$.
This finding concludes that 
the maximum shear stress that the ionic atmosphere can support is proportional to 
$\lambda_D^{-3}$, where
$\dot{\gamma}$, $\lambda_D$ and $\tau_D=\lambda_D^2/D$ are, respectively, the shear 
rate, the Debye screening length and the Debye relaxation time with $D$ being the 
relative diffusivity at the infinite dilution limit of the electrolyte.

\end{abstract}



\maketitle

\section{Introduction}
It is deeply recognized that the long-ranged Coulomb interaction between ions gives
rise to the viscosity enhancement of electrolyte solutions~\cite{harned,kunz,eyring,stokes}.
When a velocity gradient in the solution will deform an ionic atmosphere 
(a cloud of couterions around an ion), which would otherwise possesses a spherical 
symmetry, electrostatic forces 
as well as thermal motion tend to restore the atmosphere to its original form. 
Because of a finite relaxation time of the ionic atmosphere, these two effects 
can be balanced in a stationary state. 
This effect is closely related to the relaxation effect in conductance,
and also it is sometimes called the ``electroviscous'' effect by analogy with
the electroviscous effects found in charged colloidal suspensions~\cite{eyring}.
In the case of very low concentration, such an ionic effect is proportional to the square
root of the concentration, as was found many years ago by Falkenhagen~\cite{falkenhagen} 
and Onsager and Fuoss~\cite{onsager-fuoss,onsager-kim} theoretically,
and by Jones and Dole experimentally~\cite{jones-dole}.

The concentration dependence of the transport properties
(such as the conductivity and the viscosity)
of electrolyte and polyelectrolyte solutions still offer significant challenges.
Today, there exists a large number of theoretical and numerical studies 
that take into account realistic structural effects of ions and polyions.
Among them, one of the most celebrated is the mode coupling theory combined with 
accurate pair correlation functions of liquids~\cite{attard}, which has recently 
made considerable progress 
in this area~\cite{cohen,miyazaki,chandra,chandra2,bagchi}.
With regard to the study of viscosity, however, most of the literature has 
concentrated on the intrinsic viscosity.

To the best of our knowledge, nonlinear responses of electrolyte solutions in 
velocity field
remains largely unexplored so far, even in the infinite dilution limit. 
The aim of this paper is therefore to study how a strong shear flow modifies
the ``electroviscous effect'' mentioned above.
Because analytic calculations of general cases are impossible, we will
consider in this paper a binary electrolyte solution in very low 
concentration, for which analytic expressions of various statistical quantities are available.
Our formulation is in principle applicable over an unrestricted range of field
intensities as far as the local-equilibrium condition is not severely violated.
Although the present work is limited to the specific simple system 
and thus might be less noticeable in the practical point of view, 
we expect that this work may prompt further studies towards better understandings 
of rheological properties in more important systems such as concentrated electrolyte
solutions, polyelectrolyte 
solutions~\cite{jiang} and charged colloidal suspensions~\cite{zaman,menjivar}
under shear.

This paper is organized as follows.
In Sec. II, we set up Langevin equations for fluctuating 
concentration of ions combined with the free energy functional
which accounts for electrostatic interactions within Debye-H\"{u}ckel
approximation. 
Our model is directly employed from that in critical fluid binary 
mixtures~\cite{onuki-book,kawasaki}.
The linear response is first studied in Sec. III.
We show a smart derivation of Falkenhagen-Onsager-Fuoss limiting 
law~\cite{falkenhagen,onsager-fuoss,onsager-kim} for binary 
electrolyte solutions.
In Sec IV, an anisotropic structure factor of charge density
and the excess viscosity are explicitly obtained as a function of
the shear rate.
The limitation and the experimental relevancy of our results are
discussed in Sec V, followed by our brief summary in the last section.

\section{Model Equations}

\subsection{Free Energy}
Consider an electrolyte solution consisting of only two kinds of
ions dissolved in a continuum solvent of dielectric constant $\epsilon$, 
each of which carries charges $Ze$ and $-e$
($e$ is the elementally charge).
The average number density of the two ion species are written as
$\bra n_+ \ket=\bar{n}$ and $\bra n_{-}\ket=Z\bar{n}$, respectively.
Then the total ion number density is given by $n_{tot}=(Z+1)\bar{n}$.
The free energy of this system is composed of the entropic
and electrostatic contributions:
\begin{eqnarray}
 F &=\int \left[f+\frac{1}{2}\int \rho_c\phi\right]d{\bf r}.
 \label{F}
\end{eqnarray}
The first term of the right-hand side of Eq. (\ref{F}) is the entropic
contribution of ideal mixing given by~\cite{onuki-book}
\begin{eqnarray}
 f &=k_BT \sum_{\al}\left[n_{\al}\ln(n_{\al}a^3)-n_{\al}\right],
 \label{f}
\end{eqnarray}
where $n_{\al}({\bf r})$ is the number density of the ion species ${\al}$, 
$T$ the temperature, $k_B$ the Boltzmann constant, and the two kinds of ions
have a common radius $a$.
The charge density is thus given by 
$\rho_c({\bf r})=Zen_{+}({\bf r})-en_{-}({\bf r})=eZ\bar{n}\psi({\bf r})$,
where we have introduced the variable $\psi$ as the order parameter of 
the charge density fluctuation.
The second term in the free energy $F$ is the electrostatic contribution. 
The electrostatic potential satisfies the Poisson equation 
\begin{eqnarray}
 \nabla^2\phi({\bf r},t) &=-\frac{4\pi}{\epsilon}\rho_c({\bf r},t).
 \label{phi}
\end{eqnarray}
Note that the time dependence of the electrostatic potential 
entirely comes from slow variations of the charge density 
$\rho_c({\bf r},t)$ due to relative diffusion between
oppositely charged molecules.
Introducing the Coulomb operator $v({\bf r})$ which satisfies
$\nabla^2v({\bf r})=-4\pi\delta({\bf r})$, we can write the 
formal solution of Eq. (\ref{phi}) as
\begin{eqnarray}
 \phi({\bf r},t) &=\frac{1}{\epsilon}\int d{\bf r}'
	v({\bf r}-{\bf r}')\rho_c({\bf r}',t).
 \label{phi2a}
\end{eqnarray}
Assuming the total number density of ions is nearly constant everywhere:
$n_{+}({\bf r})+n_{-}({\bf r})\cong n_{tot}$,
Eq. (\ref{f}) can be expanded in powers of $\psi$. 
Up to the quadratic order, we find
$f \cong\psi^2/2\chi$,
where $\chi=(Z+1)/(Z\bar{n})$ is the osmotic compressibility.
Substituting these equations into Eq. (\ref{F}), we obtain
the free energy relevant to our purpose in the form
\begin{eqnarray}
 \beta F &=\int d{\bf r}\frac{1}{2\chi}\psi^2
	+\frac{Z^2\ell_B\bar{n}^2}{2}\int d{\bf r}\int d{\bf r}'
	\psi({\bf r})v({\bf r}-{\bf r}')\psi({\bf r}'),
 \label{f2}
\end{eqnarray}
where $\beta=1/k_BT$ and the Bjerrum length $\ell_B=e^2/\epsilon k_BT$ is defined as the
length at which the Coulomb energy between two elementary 
charges becomes comparable to thermal energy $k_BT$.

\subsection{Dynamic Equations}
To describe the dynamics of the charge fluctuation $\delta\rho_c=Ze\bar{n}\delta\psi$
and to calculate the shear viscosity under steady shear flow,
we start with the following generalized Langevin equations (see also Appendix)
\begin{eqnarray}
 \frac{\partial}{\partial t}\psi &= -{\bf v}\cdot\nabla\psi
 -L\nabla^2\frac{\delta}{\delta\psi}(\beta F)
	+\theta,
 \label{eqm_psi}\\
 \bar{\rho}\frac{\partial}{\partial t}{\bf v} &=
 -\nabla p-k_BT\psi\nabla\frac{\delta}{\delta\psi}
	(\beta F)+{\eta}_0\nabla^2{\bf v}+{\bf f},
 \label{eqm_v}
\end{eqnarray}
where the generalized diffusion equation for $\psi$ is 
now coupled with the velocity field of the fluid ${\bf v}$
through the reversible term $-{\bf v}\cdot\nabla\psi$.
The pressure $p$ in Eq. (\ref{eqm_v}) is determined so as to satisfy the incompressibility
condition $\nabla\cdot{\bf v}=0$.
The Gauss-Markov thermal noise source $\theta({\bf r},t)$
and ${\bf f}({\bf r},t)$ are related to the kinetic coefficients
$L$ and the zero-shear viscosity ${\eta}_0$ via the usual
fluctuation-dissipation relations given by
\begin{eqnarray}
 \bra\theta({\bf r},t)\theta({\bf r}',t')\ket &=
 -2L\nabla^2\delta({\bf r}-{\bf r}')\delta(t-t'),
 \label{fdt1}\\
 \bra f_i({\bf r},t)f_j({\bf r}',t')\ket &=
 -2{\eta}_0k_BT\nabla^2\delta({\bf r}-{\bf r}')\delta(t-t')
	\delta_{ij}.
 \label{fdt2}
\end{eqnarray}
The second term on the right-hand side of Eq. (\ref{eqm_v}) 
represents the force exerted to the fluid by the deviation 
of $\psi$ from its equilibrium value $\bra\psi\ket=0$.
This term should be incorporated in Eq. (\ref{eqm_v}) as the
counterpart of the reversible mode-coupling term 
$-{\bf v}\cdot\nabla\psi$ in Eq. (\ref{eqm_psi}) in order to
ensure that the equilibrium distribution of $\psi$ and ${\bf v}$
is given by the Boltzmann distribution
$\exp(-F/k_BT-\frac{\bar{\rho}}{2}\int d{\bf r}{\bf v}^2/k_BT)$.
Neglecting the random force, we can rewrite Eq. (\ref{eqm_v})
in the form which suggests the momentum conservation
\begin{eqnarray}
 \frac{\partial}{\partial t}(\bar{\rho}{\bf v})+
	\nabla\cdot{\bf \Pi} &=0,
 \label{eqm_v2}
\end{eqnarray}
where the stress tensor is written as
\begin{eqnarray}
 \Pi_{ij} &=\left(p+\frac{k_BT}{2\chi}\psi^2\right)\delta_{ij}-
	\frac{\epsilon}{4\pi}\partial_i\phi\partial_j\phi-{\eta}_0
		\bigl(\partial_iv_j+\partial_jv_i\bigr).
 \label{pi3}
\end{eqnarray}
The second term $-(\epsilon/4\pi)\partial_i\phi\partial_j\phi$ comes from
the long-ranged interionic Coulomb interactions and is nothing but 
the Maxwell stress tensor~\cite{landau}.
When the average velocity field $\bra{\bf v}\ket=\dot{\gamma}
y{\bf e}_x$ is imposed in this system, the corresponding shear stress 
is thus given by
\begin{eqnarray}
 \sigma_{xy} &=-\pbra\Pi_{xy}\pket
 = {\eta}_0\dot{\gamma}+\frac{\epsilon}{4\pi}
	\pbra\frac{\partial\phi}{\partial x}\frac{\partial\phi}
		{\partial y}\pket.
 \label{sg1}
\end{eqnarray}
Note that the second term in this equation gives rise to
the electroviscous effect in a sheared electrolyte solution.

\section{Linear Response Theory: The Limiting Law}
In this section, we shall discuss the electrostatic contribution
to the viscosity within the linear response theory.
Because we are interested in the macroscopically 
homogeneous state $(\bra\psi\ket=0)$ far from its critical point,
we can linearize Eq. (\ref{eqm_psi}) to obtain
\begin{eqnarray}
 \frac{\partial}{\partial t}\psi &=-{\bf u}\cdot\nabla\psi
 -D(\nabla^2+\kappa_D^2)\psi+\theta,
 \label{eqm_psi2}
\end{eqnarray}
where ${\bf u}=\bra{\bf v}\ket$, and where we have defined 
the relative diffusivity $D=L/\chi$ and 
the inverse Debye screening length $\kappa_D=1/\lambda_D$ as
\begin{eqnarray}
 \kappa_D^2 &=4\pi\ell_BZ^2\bar{n}^2\chi =
	\frac{4\pi Ze^2 n_{tot}}{\epsilon k_BT}.
 \label{kappa}
\end{eqnarray}
In the absence of the macroscopic flow ${\bf u}=0$, Eq. (\ref{eqm_psi2})
can be easily solved to yield the dynamic structure factor 
\begin{eqnarray}
 S(k,t) &=\frac{\bra\psi_{{\bf k}}(t)\psi_{{\bf k}'}(0)\ket}
	{(2\pi)^3\delta({\bf k}+{\bf k}')}
 \ =\ \frac{\chi k^2}{k^2+\kappa_D^2}\exp\left[-D(k^2+\kappa_D^2)t\right].
 \label{skt_psi}
\end{eqnarray}
Here, the Fourier transform of $\psi$ is defined by
\begin{eqnarray}
 \psi({\bf r},t) &=\int\frac{d^3{\bf k}}{(2\pi)^3}
	\exp\left(i{\bf k}\cdot{\bf r}\right)
	\psi_{{\bf k}}(t).
 \label{ft_psi}
\end{eqnarray}
The intrinsic viscosity including the electrostatic contribution 
${\eta}$ can be calculated by making use of the Green-Kubo 
formula~\cite{onuki-book}:
\begin{eqnarray}
 {\eta} &={\eta}_0 +\frac{1}{k_BT}\int_0^{\infty}dt
	\int d{\bf r}\pbra\hat{\Pi}_{xy}({\bf r},t)
		\hat{\Pi}_{xy}({\bf 0},0)\pket,
 \label{g-k}
\end{eqnarray}
where the electrostatic shear stress is given by 
$\hat{\Pi}_{xy}=\epsilon/4\pi(\partial\phi/\partial x)
(\partial\phi/\partial y)$.
Within the Gaussian approximation, the four-body correlation 
between $\phi$ appearing in the right-hand side of Eq. (\ref{g-k}) 
can be decoupled into the product of the pair correlation of $\phi$.
Noting that the Fourier transform of Eq. (\ref{phi}) gives
$\phi_{{\bf k}}(t)=4\pi Ze\bar{n}(\epsilon k^2)^{-1}\psi_{{\bf k}}(t)$ and
using (\ref{skt_psi}), we can readily integrate over time in Eq. (\ref{g-k})
to find
\begin{eqnarray}
 {\eta} &={\eta}_0+\frac{k_BT}{D}\kappa_D^4
	\int\frac{d^3{\bf k}}{(2\pi)^3}\frac{k_x^2k_y^2}{k^4(k^2+\kappa_D^2)^3}
 \label{g-k2}
\end{eqnarray}
Performing the integration over ${\bf k}$, we finally obtain 
${\eta}={\eta}_0+c_1k_BT\kappa_D/D$, where the numerical constant $c_1$
is given by 
\begin{eqnarray}
 c_1 &=\frac{1}{(2\pi)^3}\int_0^{2\pi}\sin^2\varphi\cos^2\varphi d\varphi
 \int_0^{\pi}\sin^5\theta d\theta\int_0^{\infty}\frac{x^2 dx}{(1+x^2)^3}
 =\frac{1}{480\pi}.
 \label{cons}
\end{eqnarray}
The relative diffusivity $D$ in Eq. (\ref{eqm_psi2}) has been
originally introduced in the context of ``collective'' meaning. 
In the limit of the infinite dilution, however, $D$ coincides with
the self diffusivity of a tagged ion, $D_s$, which is related to 
the mobility of the tagged ion $\mu$ via the Einstein relation $D_s=\mu k_BT$.
Therefore, by defining the friction constant of a tagged ion 
as $\gamma=\mu^{-1}$, one can recover the limiting law originally derived by  
Falkenhagen for binary electrolytes~\cite{falkenhagen} (and later rederived by  
Fuoss and Onsager for more general multicomponent cases
~\cite{onsager-fuoss,onsager-kim}), which is written as
\begin{eqnarray}
 \Delta\eta_0&={\eta}-{\eta}_0=\frac{\kappa_D\gamma}{480\pi}.
 \label{eta2}
\end{eqnarray}
This simple scaling concludes the well-known fact that the viscosity 
enhancement is proportional to the square root of the electrolyte
concentration. 
A similar derivation of Eq. (\ref{eta2}) 
can be also found in the recent literature~\cite{chandra,chandra2}.

\section{Nonlinear Effects} 
\begin{figure}
\begin{center}
\includegraphics[width=0.707\linewidth]{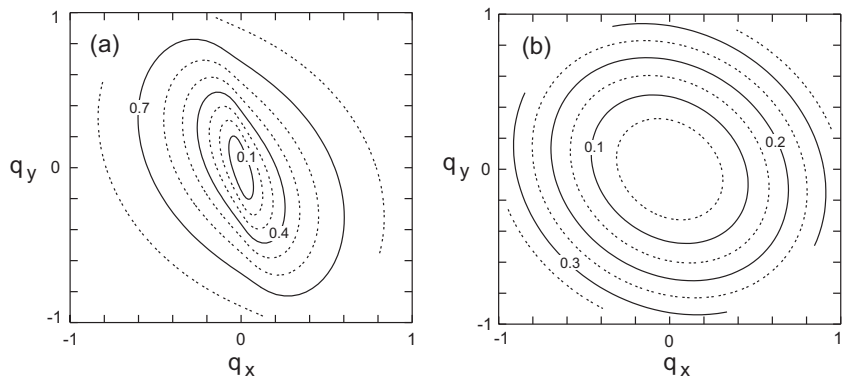}
\end{center}
\caption{\label{fig0}
Contour plots in the shear plane of the steady-state structure factor 
$\tilde{S}({\bf q})$ for shear rates (a) $\tau_D\dot{\gamma}=5.0$ and 
(b) $\tau_D\dot{\gamma}=0.5$.}
\end{figure}

Let us proceed to investigate how the electroviscous effect is
modified in the presence of steady shear flow.
In the case of simple shear ${\bf u}=\dot{\gamma}y{\bf e}_x$, 
${\bf e}_x$ being the unit vector along the $x$ axis, the diffusion 
equation (\ref{eqm_psi2}) in the Fourier space becomes
\begin{eqnarray}
 \left[\frac{\partial}{\partial t}-\dot{\gamma}k_x
	\frac{\partial}{\partial k_y}\right]\psi_{{\bf k}}(t)
	&=-D(k^2+\kappa_D^2)\psi_{{\bf k}}(t)+\theta_{{\bf k}}(t).
 \label{meq_psi3}
\end{eqnarray}
Applying the standard manipulation described somewhere~\cite{dufty},
we find the solution of Eq. (\ref{meq_psi3}) in the form
\begin{eqnarray}
 \psi_{{\bf k}}(t) &=\int_0^{\infty}\theta_{{\bf k}(s)}(t-s)
	G({\bf k},s)ds,
 \label{sol01}
\end{eqnarray}
where ${\bf k}(t)$ is the time-dependent wave vector defined by
${\bf k}(t)={\bf k}+\dot{\gamma}k_xt{\bf e}_y$,
and the Green's function is given by
\begin{eqnarray}
 G({\bf k},t) &=\exp\left[-D \int_0^t ds 
	({\bf k}^2(s)+\kappa_D^2)\right].
 \label{green}
\end{eqnarray}
Using the correlation property Eq. (\ref{fdt1}), we find that
the steady-state structure factor is given by
\begin{eqnarray}
 S({\bf k}) &=2L \int_0^{\infty}{\bf k}^2(s)G^2({\bf k},s)ds.
 \label{sk_s}
\end{eqnarray}
If we introduce the dimensionless variables ${\bf q}=(D/\dot{\gamma})^{1/2}{\bf k}$
and $l=\dot{\gamma}s$, Eq. (\ref{sk_s}) can be rewritten as
$S({\bf k})=\chi\tilde{S}({\bf q})$, where $\tilde{S}({\bf q})$ is 
explicitly given by
\begin{eqnarray}
 \tilde{S}({\bf q}) &= 1-\frac{2}{\tau_D\dot{\gamma}}\int_0^{\infty}dl
 \exp\left[-2\left\{\left(q^2+\frac{1}{\tau_D\dot{\gamma}}\right)l+
		q_xq_yl^2+\frac{1}{3}q_x^2l^3\right\}\right].
 \label{sq_s}
\end{eqnarray}
The Debye relaxation time $\tau_D$ defined by $\tau_D=\lambda_D^2/D=
\epsilon k_BT/(4\pi Ze^2n_{tot}D)$ represents the time scale at which the deformed 
ionic atmosphere around an ion recovers its equilibrium spherical form.
The dimensionless quantity $\tau_D\dot{\gamma}$ is therefore regarded as
the control parameter which measures the relative strength of the
shear flow (external perturbation) to the stability inherent to this
system.
Numerically evaluated $\tilde{S}({\bf q})$ for two different values of
$\tau_D\dot{\gamma}$ is displayed in Fig. \ref{fig0}.
For $\tau_D\dot{\gamma}>1$, the structure factor is considerably deformed by
the shear, and rapidly reaches the unity as $q_x$ increases, reflecting
the enhanced homogeneization along the flow direction by the shear.

These dramatic changes of $S({\bf k})$ due to shear should also affect the ionic
contribution to the viscosity.
According to Eq. (\ref{sg1}), the excess shear stress is given by
\begin{eqnarray}
 \sigma_{xy}(\dot{\gamma}) &=\frac{\epsilon}{4\pi}\left(
	\frac{4\pi Ze\bar{n}}{\epsilon}\right)^2
	\int\frac{d^3{\bf k}}{(2\pi)^3}\frac{k_xk_y}{k^4}S({\bf k}).
 \label{stress1}
\end{eqnarray}
As is obvious from the spherical symmetry, $\sigma_{xy}(\dot{\gamma}=0)$ is zero. 
However, to avoid apparent divergence of the integrand in Eq. (\ref{stress1}) in the limit of 
$s=0$, we hereafter consider the quantity
$\Delta\sigma_{xy}(\dot{\gamma})=\sigma_{xy}(\dot{\gamma})-\sigma_{xy}(0)$.
Introducing the polar coordinate and integrating over $k=|{\bf k}|$,
we have
\begin{eqnarray}
 \Delta\sigma_{xy}(\dot{\gamma}) &=\frac{1}{8}
  \frac{k_BT\kappa_D^2}{(2\pi)^{5/2}D^{1/2}}\int_0^{2\pi}d\varphi\int_0^{\pi}d\theta
	\,\sin\theta\,a(\theta,\varphi) \nonumber \\
 &\times \int_0^{\infty}ds\,e^{-D\kappa_D^2 s}
	\left[\frac{1+2a(\theta,\varphi)\dot{\gamma}s+b(\theta,\varphi)\dot{\gamma}^2s^2}
	{(s+a(\theta,\varphi)\dot{\gamma}s^2+\frac{1}{3}b(\theta,\varphi)
		\dot{\gamma}^2s^3)^{3/2}}-\frac{1}{s^{3/2}}\right].
 \label{stress3}
\end{eqnarray}
where $a(\theta,\varphi)=\sin^2\theta\cos\varphi\sin\varphi$ and 
$b(\theta,\varphi)=\sin^2\theta\cos^2\varphi$ have been introduced.
Integrating further over $s$ by parts, we finally arrive at the following
result after some algebra,
\begin{eqnarray}
 \Delta{\eta}(\dot{\gamma}) &=\frac{1}{\dot{\gamma}}\Delta\sigma_{xy}(\dot{\gamma})
	\ =\ \Delta {\eta}_0H(\tau_D\dot{\gamma}),
 \label{eta_s}
\end{eqnarray}
where $\Delta{\eta}_0=(480\pi D)^{-1}k_BT\kappa_D$ is the excess 
viscosity derived in the last section.
The scaling function $H(x)$ is written as
\begin{eqnarray}
 H(x) &=\frac{240}{(2\pi)^{3/2}}\frac{1}{x}\int_0^{\infty}
	e^{-2u^2}J(xu^2)du,
 \label{omega2} 
\end{eqnarray}
where 
\begin{eqnarray}
 J(y) &=\int_0^{2\pi}d\varphi\int_0^{\pi}d\theta
	\,\sin\theta\,a(\theta,\varphi)\left[
	1-\frac{1}{(1+a(\theta,\varphi)y+\frac{1}{3}b(\theta,\varphi)y^2)^{1/2}}\right].
 \label{jy}
\end{eqnarray}
By expanding $J(xu^2)$ in powers of $x$, one can easily 
check $H(x)=1-b_1x^2+\cdots$ for $x \ll 1$, where $b_1=15/128$. 
The present theory therefore correctly reduces to the Falkenhagen-Onsager-Fuoss
expression of the excess viscosity in the zero-shear limit.
For small $\tau_D\dot{\gamma}$, the excess viscosity is the analytic 
function of $\dot{\gamma}$ and exhibits a weak shear-thinning given by
\begin{eqnarray} 
 \Delta{\eta}(\dot{\gamma})&=\Delta{\eta}_0\left[
	1-\frac{15}{128}(\tau_D\dot{\gamma})^2+\cdots\right].
 \label{expansion}
\end{eqnarray} 

\begin{figure}
\begin{center}
\includegraphics[width=0.816\linewidth]{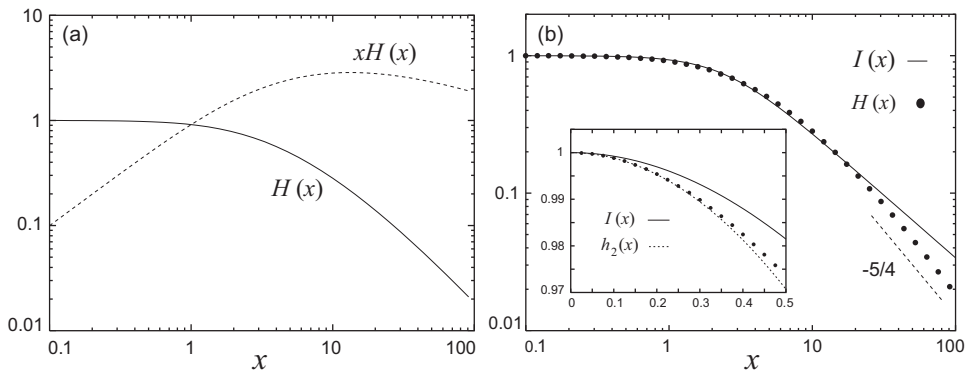}
\end{center}
\caption{\label{fig1}
(a) Plots of numerically evaluated scaling functions $H(x)$ and $xH(x)$.
(b) Comparison between (numerically evaluated) $H(x)$ and the fitting function
$I(x)$ [Eq. (\ref{ix})].}

\end{figure}

\section{Discussion}
The numerical evaluation of the scaling function $H(x)$ is shown in Fig. \ref{fig1}.
As can be seen in Fig. \ref{fig1} (a), the excess viscosity $\Delta\eta$ rapidly decreases 
much below the intrinsic value $\Delta\eta_0$ as the shear rate exceeds 
$\dot{\gamma}_c=\tau_D^{-1}$.
The overall behavior of $H(x)$ up to $x\sim20$ are
fairly well fitted by the empirical function $I(x)$ of the form
\begin{eqnarray}
 I(x)&=\frac{1}{\left[1+\lambda x^2\right]^{\alpha}},
 \label{ix}
\end{eqnarray}
where $\lambda\cong0.17$ and $\alpha\cong0.46$ in Fig. \ref{fig1} (b).
For small enough $x$, however, $h_2(x)=1-b_1x^2$ (the Taylor expansion 
of $H(x)$ around $x=0$ which we have obtained in the last section) actually
fits better to the numerical date of $H(x)$ than $I(x)$, as shown in the inset 
of Fig. \ref{fig1} (b).   
Roughly speaking, therefore, the excess viscosity behaves as 
$\Delta\eta\sim\dot{\gamma}^{-2\alpha}$ for 
$1<\tau_D\dot{\gamma}<20$, 
and even more strongly decreases as $\Delta\eta\sim\dot{\gamma}^{-5/4}$ 
for larger $\tau_D\dot{\gamma}$.

Correspondingly, the excess shear stress $\Delta\sigma_{xy}$ saturates into its 
maximum value $\Delta\sigma_m \sim \Delta\eta_0\dot{\gamma}_c\sim c_1k_BT/\lambda_D^3$ 
at $\tau_D\dot{\gamma}\sim 10$, and then slowly decreases for larger $\dot{\gamma}$.
If one considers the ``effective surface tension'' $a_s$ of an ionic atmosphere of size
$\lambda_D$ on analogy with a droplet in an immiscible fluid, one would obtain 
$a_s \sim k_BT/\lambda_D^2$ from the energy balance condition.
Assuming that the pressure difference between interior and exterior of the droplet
can be balanced by the imposed shear stress, 
$a_s/\lambda_D \sim \Delta\sigma_m$, one immediately recovers the above expression of
$\Delta\sigma_m$.

Let us examine the magnitude of the predicted shear-thinning effect.
For the zero-shear viscosity at higher concentrations, the following 
empirical equation:
\begin{eqnarray}
 \Delta\eta_0(c) &= A\sqrt{c}+B c,
 \label{del_eta2}
\end{eqnarray}
provides a better description of the experimental results.
The constant $A$ was evaluated from experimental data for various salts
by plotting a suitable form of Eq. (\ref{del_eta2}), and the good agreement
with the theory has been obtained for aqueous solutions~\cite{stokes}.
However, the absolute magnitude of this contribution to the
viscosity ($A\sqrt{c}$ term) is found to be rather small (as can be seen 
in the large number (480$\pi$) in the denominator of Eq. (\ref{eta2})).
Usually, more significant contribution comes from the ion-solvent interaction,
which is represented by the $B c$ term in Eq. (\ref{del_eta2}), but 
there is no satisfactory
theoretical explanation of the coefficient $B$ in Eq. (\ref{del_eta2}) so far.
The present study suggests that, at strong shear as $\tau_D\dot{\gamma}\sim10$,
the coefficient $A$ decreases to 20\% of its original value.
Therefore, while a direct experimental observation of the shear-thinning would
be hard because of the vanishingly small value of $A$, the modified fitting 
function $\Delta\eta_0/\eta_0\cong 1+Bc$, which tells the linear dependence of
the viscosity upon the concentration, might be best fitted to experimental
data under strong shear conditions (provided that the linear concentration 
dependence arising from the ion-solvent interaction is not significantly changed
by the shear).
Although indirect, this may serve a possible experimental verification of the 
shear-thinning effect in future experiments.

At very large shear $\tau_D\dot{\gamma}\gg 1$, the ionic atmosphere is no longer
ellipsoidal because it is dramatically stretched by the shear, and dissipates
thermally. 
The considerably deformed atmosphere may not support the shear stress any more,
and $\Delta\sigma_{xy}$ itself is thus decreased.
If we use typical values of the diffusion constant $D\sim 10^{-5}$ cm$^2$sec$^{-1}$
and the Debye screening length $\lambda_D\sim 3\times10^{-7}$ cm for 0.01 M solutions
of strong 1:1 electrolytes (like NaCl in water), 
we find $\dot{\gamma}_c\sim 10^8$ sec$^{-1}$, which is unrealistically large.
This simple estimation suggests that, unfortunately, even the strong shear condition 
$\tau_D\dot{\gamma}\sim 1$ might be difficult to be realized in ordinary experimental
setup.
In other words, the ionic contribution to the viscosity may not be significantly 
influenced by the shear in most realistic conditions.

Experimentally, however, a near-critical ionic fluid, in which the charge density 
fluctuation decays slowly, can be prepared by carefully choosing materials 
and by controlling the temperature~\cite{pitzer,weingartner,narayanan,koulinskii}.
In that case, the correlation decays as $\exp(-D_Rk^2t)$,
where the (renormalized) diffusion constant may be given
by Kawasaki-Stokes's law $D_R\sim k_BT/6\pi{\eta}\xi$ with $\xi$ being the
correlation length of the critical fluctuations~\cite{onuki-book,kawasaki}.
Although a rather crude argument, we naively expect $\dot{\gamma}_c \sim 3\times 10^4$
sec$^{-1}$, which is still very large but is much lower than that for an ordinary electrolyte
solution, as well as a relatively large magnitude of the electroviscous effect on
the order of $\Delta\eta_0/\eta_0\cong\frac{1}{80}\kappa_D\xi \sim 0.1$, if we assume 
$\eta_0\sim 40$ cP, $\xi\sim 10$ nm, $\lambda_D\sim 1$nm for $T=343$ K, which are the
typical value set for (Bu$_4$NPic)-tridecane-1-ol.
(This model system shows liquid-liquid phase separation with an upper consolute
point near 342K, expected to be driven by Coulombic interaction~\cite{weingartner}.)
Needless to say, in order for a more accurate treatment of an ionic fluid near a demixing 
or a gas-liquid transition, 
we have to take into consideration the density gradient term 
$(\nabla \psi)^2$ in $F$ which accounts for
the spatial heterogeneity of critical fluctuations (and becomes more
and more important as the system approaches the critical point).
Our rough estimation therefore might be appropriate only in pre-stages of a critical
regime. 
In this connection, the free energy Eq.(\ref{f2}) including the gradient term is 
mathematically equivalent to that of diblock copolymer systems
derived by Ohta and Kawasaki~\cite{ohta}, in which the long-range interaction
arising from the connectivity of different chemical sequences in a polymer chain
plays a crucial role.
Its close similarity to charged fluids has also been pointed out in the literature~\cite{ohta}.
Near-critical properties of that system under shear in one phase region was also studied 
by Onuki based on the same dynamic equations employed here~\cite{onuki}.
On the other hand, in molecular dynamics simulations, accessible shear rates are
much larger than those in real experiments, and the strong shear condition will be
easily realized~\cite{todd,bair}.
Then the pronounced shear-thinning behavior predicted here would be more likely to
be checked by computer simulations.

The increase and saturation of ionic conductivity of electrolyte 
solutions at very high applied voltage is known as 
Wien effect~\cite{harned,onsager-kim2,netz}.
This increase has been shown in general to arise from two different 
effects, namely, a destruction of the ionic atmosphere and a modification
of the dissociation kinetics~\cite{onsager-kim2}.
In strong electrolytes the former effect can dominate over the latter.
It is instructive to remark that the similar mechanism works in the 
system we studied here, that is, the destruction of the ionic atmosphere by
strong shear renders the increase and saturation of the reciprocal of the 
viscosity, {\it i.e.}, the fluidity.

\section{Conclusion}
We now summarize our results.
Effects of shear flow on the viscosity of electrolyte solutions are studied
theoretically within the framework of fluctuating hydrodynamics.
Our calculations confirm the Falkenhagen-Onsager-Fuoss limiting law that 
the ionic effects
to the viscosity is proportional to the square root of the concentration of
the electrolytes in the dilute and the zero-shear limit.
We extend this classical result for finite shear by deriving the analytic 
expression of the excess nonlinear shear viscosity.
The electrostatic contribution to the viscosity is drastically reduced under
the strong shear condition $\tau_D\dot{\gamma}>1$, which however might be difficult 
to access experimentally.
Because the use of Debye-H\"{u}ckel theory here limits the applicability of our 
treatment to low concentration solutions, the viscoelasticity of concentrated 
electrolyte solutions under shear is an important remaining problem.
Studies in this direction using Brownian Dynamics simulation is currently in progress.  
We hope that this work will prompt further studies on the nonlinear transport 
theory of electrolyte solutions in large velocity fields,
which might constitute the counterpart of the conductivity studies in large driving fields.

\ack{
The author would like to thank Y. Murayama and M. Sano for valuable discussions.
This work was supported by Grant-in-Aid for JSPS
Fellows for Young Scientists, from Ministry of Education,
Culture, Sports, Science, and Technology, Japan.
}

\appendix
\section{Derivation of the dynamic equations}
In this Appendix, we describe the derivation of the dynamical equations 
(\ref{eqm_psi}) and (\ref{eqm_v}) based on a two-fluid model for electrolyte solutions.
We will specifically consider strongly dissociative salt 
solutions such as NaCl and KCl in water.
Such a solution should be modelled as a three-fluid model, namely, 
as a mixture of dissolved positive ions, negative ions and the solvent.
For simplicity, however, we will start from the two-fluid model for positive and 
negative ions, while regarding the solvent as the structureless 
background with a static dielectric constant $\epsilon$.
A number of approximations and simplifications involved at this stage 
are to be critically checked in future studies.
(The present formulation is convenient to directly obtain a mutual 
diffusion equation for electrolyte solutions coupled with
the Navier-Stokes equation. 
However, in order to obtain a diffusion equation for each ion 
species with vanishing electrostatic interactions, we also have to take into account 
hydrodynamic drag forces between ions and solvent molecules, which are not 
explicitly included in the present formulation.
A more precise and systematic treatment for diffusion equations can be
found, for example, in Ref.~\cite{dufreche}.)  

The number conservation law for the ionic species $\al=\pm$ is written as
\begin{eqnarray}
 \pder{n_{\al}}{t}+\na\cdot(n_{\al}{\bf v}_{\al}) &= 0,
 \label{app2:eq1}
\end{eqnarray}
while the total density $n_{tot}=n_++n_-$ obeys the usual continuity equation
\begin{eqnarray}
 \pder{n_{tot}}{t}+\na\cdot(n_{tot}{\bf v}) &= 0,
 \label{app2:eq2}
\end{eqnarray}
where the average velocity ${\bf v}$ is defined by
\begin{eqnarray}
 n_{tot}{\bf v} &= n_+{\bf v}_++n_-{\bf v}_-.
 \label{app2:eq3}
\end{eqnarray}
Assuming that a fluid under consideration shows a vanishingly small
compressibility in long-time scales, we set $\na\cdot{\bf v}=0$ 
(the incompressibility condition), which implies from Eq. (\ref{app2:eq2})
that the total number density is constant,
$n_{tot}\cong\bar{n}_{tot}=(Z+1)\bar{n}$, for a sufficiently slow fluid 
motion where $\pa_t n_{tot}$ can be neglected.
Introducing a relative velocity between the positive and negative ions by
${\bf u}={\bf v}_+-{\bf v}_-$, 
we can write the velocities of the two ion species as
\begin{equation}
 {\bf v}_+ ={\bf v}+\frac{n_-}{n_{tot}}{\bf u},\qquad 
 {\mbox{and}} \qquad 
	{\bf v}_- ={\bf v}-\frac{n_+}{n_{tot}}{\bf u}.
 \label{app2:eq5}
\end{equation}
Since we have defined a scaled charge density fluctuation as
$Ze\bar{n}\psi({\bf r})=Zen_+({\bf r})-en_-({\bf r})$, 
we find from Eqs. (\ref{app2:eq1}) and (\ref{app2:eq5}) the equations for $\psi$:
\begin{eqnarray}
 \pder{\psi}{t}+\na\cdot(\psi{\bf v}) &=
	-\na\cdot\left[\frac{1}{Z}\frac{n_+n_-}{{\bar{n}}^2}
		{\bf u}\right].
 \label{app2:eq7}
\end{eqnarray}
On the other hand, the equations of motions for the two ions species may be given by
\begin{eqnarray}
 mn_+\pder{{\bf v}_+}{t} &= -n_+\na\mu_++\eta_+\na^2{\bf v}_+-\zeta{\bf u},
 \label{app2:eq8a}\\
 mn_-\pder{{\bf v}_-}{t} &= -n_-\na\mu_-+\eta_-\na^2{\bf v}_-+\zeta{\bf u},
 \label{app2:eq8b}
\end{eqnarray}
where $m$ is the equal mass for two ion species, $\eta_{\al}$ is the viscosity, 
$\mu_{\al}$ is the appropriately defined chemical potential, and $\zeta$ is the 
friction constant between the two ion species. 
Note that $\zeta$ may be given in terms of the friction constant of the
each ion species $\zeta_{\al}$ by $\zeta^{-1}=\zeta_{+}^{-1}+\zeta_{-}^{-1}$.
Subtracting Eq. (\ref{app2:eq8b}) from Eq. (\ref{app2:eq8a}), we obtain
the equation for the relative velocity ${\bf u}$ as
\begin{eqnarray}
 \pder{\bf u}{t} &= -\frac{1}{m}\na(\mu_+-\mu_-)
	-\frac{\zeta}{m}\left(\frac{1}{n_+}+\frac{1}{n_-}\right){\bf u},
 \label{app2:eq9}
\end{eqnarray}
where the viscosity terms have been neglected.
Since we are concerned with very slow motion whose characteristic frequencies are
much lower than $\zeta(1/n_{+}+1/n_{-})$, we can set $\pa_t{\bf u}=0$
in Eq. (\ref{app2:eq9}) to obtain
\begin{eqnarray}
 {\bf u} &\cong -\left(\frac{1}{n_+}+\frac{1}{n_-}\right)^{-1}\frac{1}{\zeta}
	\na(\mu_+-\mu_-).
 \label{app2:eq10}
\end{eqnarray}
Here the chemical potential difference between two components are connected to
the free energy as 
\begin{eqnarray}
 \mu_+-\mu_- &= \frac{\delta F}{\delta n_+}-\frac{\delta F}{\delta n_-}=
	{\rm const} +\chi\frac{\delta F}{\delta \psi},
 \label{app2:eq11}
\end{eqnarray}
where $F$ is the free-energy functional in our model, and
$\chi=(Z+1)/(Z\bar{n})$.
In differentiating $F$ with respect to $n_+$ ($n_-$), we fix
$n_-$ ($n_+$) in Eq. (\ref{app2:eq11}).
This relation can be directly checked by using the explicit form of
$F$ given in the main text.
Substituting Eq. (\ref{app2:eq10}) into Eq. (\ref{app2:eq7}) with
Eq. (\ref{app2:eq11}), we obtain
\begin{eqnarray}
 \pder{\psi}{t}+\na\cdot(\psi{\bf v}) &=
	L\na^2\frac{\delta}{\delta \psi}(\beta F),
 \label{app2:eq12}
\end{eqnarray}
where $\beta=1/k_BT$ and the kinetic coefficient
\begin{eqnarray}
 L &= \frac{k_BT}{Z}\frac{\chi}{\zeta}\frac{(\bar{n}_+\bar{n}_-)^2}{{\bar{n}}^2
	\bar{n}_{tot}} = \frac{k_BT}{\zeta},
 \label{app2:eq13}
\end{eqnarray}
is independent of $\psi$.
On the other hand, by adding Eq. (\ref{app2:eq8a}) and (\ref{app2:eq8b}), the equation for 
the average velocity ${\bf v}$ is found to become 
\begin{eqnarray}
 m\bar{n}_{tot}\pder{{\bf v}}{t} &= -(n_+\na\mu_++n_-\na\mu_-)+
	(\eta_{+}+\eta_{-})\na^2{\bf v},
 \label{app2:eq14}
\end{eqnarray}
where the use of Eq. (\ref{app2:eq5}) has been
made, and only fluctuations up to the linear order have been retained.
Again, by using the explicit form of the free energy functional, one 
can directly check the relation
\begin{eqnarray}
 n_+\na\mu_++n_-\na\mu_- &= \na p-\psi\na\frac{\delta F}{\delta \psi},
 \label{app2:eq15}
\end{eqnarray}
where $p$ can be regarded as the hydrostatic pressure of the fluid.
At a final step, adding random thermal noise sources $\theta({\bf r},t)$
and ${\bf f}({\bf r},t)$ which ensure the appropriate equilibrium 
correlations of $\psi$ and ${\bf v}$ to Eqs. (\ref{app2:eq12}) and 
(\ref{app2:eq14}), we obtain a closed set of hydrodynamic equations
for $\psi$ and ${\bf v}$
\begin{eqnarray}
 \frac{\partial}{\partial t}\psi &= -{\bf v}\cdot\nabla\psi
 -L\nabla^2\frac{\delta}{\delta\psi}(\beta F)
	+\theta,
 \label{app2:eqm_psi}\\
 \bar{\rho}\frac{\partial}{\partial t}{\bf v} &=
 -\nabla p-k_BT\psi\nabla\frac{\delta}{\delta\psi}
	(\beta F)+{\eta}_0\nabla^2{\bf v}+{\bf f},
 \label{app2:eqm_v}
\end{eqnarray}
where $\bar{\rho}=m\bar{n}_{tot}$ is the average mass density,
$\eta_0=\eta_++\eta_-$ is the total viscosity of the fluid, and where
the incompressibility condition $\na\cdot{\bf v}=0$ has been used 
in Eq. (\ref{app2:eqm_psi}).


\section*{References}
\begin{thebibliography}{99}

\bibitem{harned} Harned H S and Owen B B, 1950 {\it Physical Chemistry of Electrolytic
Solutions} 
(New York: Reinhold).

\bibitem{kunz} Barthel J M G, Krienke H and Kunz W, 1998 {\it Physical Chemistry
of Electrolyte Solutions: Modern Aspects} (New York: Springer).

\bibitem{eyring} Conway B E, 1970 {\it Physical Chemistry} vol~IXA Electrochemistry
edited by Eyring H \etal. (London: Academic Press) p~1.

\bibitem{stokes} Stokes R H and Mills R, 1965 {\it Viscosity of Electrolytes and 
Related Properties} (Oxford: Pergamon Press).

\bibitem{falkenhagen} Falkenhagen H and Vernon E L, 1932 {\it Philos. Mag.} {\bf 14}
537.

\bibitem{onsager-fuoss} Onsager L and Fuoss R, 1932 {\it J. Phys. Chem.} {\bf 36}
2689.

\bibitem{onsager-kim} Onsager L and Kim S K, 1957 {\it J. Phys. Chem.} {\bf 61}
215.

\bibitem{jones-dole} Jones G and Dole M, 1929 {\it J. Am. Chem. Soc.} {\bf 51} 2950.

\bibitem{attard} Attard P, 1993 {\it Phys. Rev. E} {\bf 48} 3604.

\bibitem{cohen} Cohen J, Priel Z and Rabin Y, 1988 {\it J. Chem. Phys.}
{\bf 88} 7111.

\bibitem{miyazaki} Miyazaki K, Bagchi B and Yethiraj A, 2004 {\it J. Chem. Phys.}
{\bf 121}, 8120.

\bibitem{chandra} Chandra A and Bagchi B, 2000 {\it J. Phys. Chem. B} {\bf 104} 9067.

\bibitem{chandra2} Chandra A and Bagchi B, 2000 {\it J. Chem. Phys.} {\bf 113} 3226.

\bibitem{bagchi} Dufreche J-F, Bernard O, Turq P, Mukherjee A and Bagchi B, 2002
{\it Phys. Rev. Lett.} {\bf 88} 095902.

\bibitem{jiang} Jiang L, Yang D. and Chen S B, 2001 {\it Macromolecules} {\bf 34}
3730.

\bibitem{zaman} Zaman A A and Moudgil, 1999 {\it J. Colloid Interface Sci.}
{\bf 212} 167.

\bibitem{menjivar} Menjivar J A and Rha C, 1983 {\it J. Chem. Phys.} {\bf 79}
953.

\bibitem{onuki-book} Onuki A 2002 {\it Phase Transition Dynamics} 
(Cambridge: Cambridge University Press).

\bibitem{kawasaki} Kawasaki K 1976 in {\it Phase Transitions and Critical Phenomena}
eds. Domb C and Green M S (New York: Academic Press) Vol.5A, p~165.

\bibitem{landau} Laudau L D and Lifshitz E M, 1984 {\it Electrodynamics of Continuous
Media} (Oxford: Pergamon Press) Vol. 8.

\bibitem{dufty} Dufty J W 1984 {\it Phys. Rev. A} {\bf 70} 31.

\bibitem{pitzer} Pitzer K S, Bischoff J L and Rosenbauer R J, 1987
{\it Chem. Phys. Lett.} {\bf 134} 60.

\bibitem{weingartner} Weingartner H, Wiegand S and Schroer W,
1991 {\it J. Chem. Phys.} {\bf 96} 848.

\bibitem{narayanan} Narayanan T and Pitzer K S, 1995 {\it J. Chem. Phys.}
{\bf 102} 8118.

\bibitem{koulinskii} Koulinskii V L, Malomuzh N P and Tolpekin V A, 1999
{\it Phys. Rev. E} {\bf 60} 6897.

\bibitem{ohta} Ohta T and Kawasaki K, 1986 {\it Macromolecules} {\bf 19} 2621.

\bibitem{onuki} Onuki A, 1987 {\it J. Chem. Phys.} {\bf 87} 3692.

\bibitem{todd} Marcelli G, Todd B D and Sadus R J, 2001 {\it Phys. Rev. E}
{\bf 63} 021204.

\bibitem{bair} Bair S, McCabe C and Cummings P T, {\it Phys. Rev. Lett.} {\bf 88}
058302.

\bibitem{onsager-kim2} Onsager L and Kim S K, 1957 {\it J. Phys. Chem.} {\bf 61}
198.

\bibitem{netz} Netz R R, 2003 {\it Europhys. Lett.} {\bf 63} 616.

\bibitem{dufreche} Dufreche J-F, Bernard O  and Turq P, 2002 {\it J. Chem. Phys.}
{\bf 116} 2085.
 
\end{thebibliography}
\end{document}